\documentclass[aps,pre,groupedaddress,onecolumn]{revtex4}
\pdfoutput=1

\newcommand{\Dc}{\mathcal{D}}
\newcommand{\Jc}{\mathcal{J}}
\newcommand{\Pc}{\mathcal{P}}
\newcommand{\Qc}{\mathcal{Q}}
\newcommand{\Hc}{\mathcal{H}}
\newcommand{\Lc}{\mathcal{L}}
\newcommand{\Sc}{\mathcal{S}}

\usepackage{ifpdf}
\ifpdf
  \usepackage[pdftex]{graphicx}
  \else
  \usepackage{graphicx}
  \usepackage{bm}
\fi

\usepackage{hyperref}
\usepackage{subfigure}
\usepackage{color}
\usepackage{graphicx}

\begin{document}
\expandafter\ifx\csname urlprefix\endcsname\relax\def\urlprefix{URL }\fi

\ifpdf
\DeclareGraphicsExtensions{.pdf, .jpg}
\else
\DeclareGraphicsExtensions{.eps, .jpg}
\fi

\title{\large
   Bohmian trajectories and the Path Integral Paradigm.\\
   Complexified Lagrangian Mechanics.  }

\large
\author{Valeriy I. Sbitnev}
\email{valery.sbitnev@gmail.com}
\affiliation{B.P.Konstantinov St.-Petersburg Nuclear Physics Institute, Russ. Ac. Sci.,
     Gatchina, Leningrad district, 188350, Russia.}

% Comments: 14 pages, 3 figures
% Subjects: Quantum Physics (quant-ph)

\date{\today}

\begin{abstract}
 David Bohm  shown that the Schr{\"o}dinger equation, that is a "visiting card" of quantum mechanics,
 can be decomposed onto two equations for real functions - action and probability density.
 The first equation is the Hamilton-Jacobi (HJ) equation, a "visiting card" of classical mechanics,
 to be modified by the Bohmian quantum potential.
 And the second is the continuity equation. The latter can be transformed to the entropy balance equation.
 The Bohmian quantum potential is transformed to two Bohmian quantum correctors.
 The first corrector modifies kinetic energy term of
 the HJ equation, and the second one modifies potential energy term.
 Unification of the quantum HJ equation and the entropy balance equation gives complexified HJ equation
 containing complex kinetic and potential terms.
 Imaginary parts of these terms have order of smallness about the Planck constant.
 The Bohmian quantum corrector is indispensable term modifying
 the Feynman's path integral by expanding coordinates and momenta to imaginary sector.


\end{abstract}

\maketitle

% Keywords: Schr{\"o}dinger equation; Bohmian quantum potential; Buhmian trajectory; Hamilton-Jacobi equation;
% Legendre's dual transformation; principle of least action; Feynman's path integral

\large

\section{\label{sec:level1}Introduction.}
 Let us suppose, a quantum automaton within a quasi-molecular maze searches an optimal path from point A to B.
 It is a navigating problem.  At solving navigation of a mobile robot, its solution boils down to consideration of appropriate
 Bellman-Hamilton-Jacobi equation~\cite{Bellman1957},~\cite{LaValle2006}.
 As for the quasi-classical mechanics problem,
 it refers to the classical Hamilton-Jacobi equation (HJ equation)~\cite{Lanczos1970}.
 The classical theory says - the optimal path results from the principle of least action, $\delta J=0$.
 The action $J$ is an integral along the path
 from A (starting at the moment $t_{\,0}$) to B (finishing at the moment
 $t_{\,1}$)
\begin{equation}\label{qp=1}
    J=\int\limits_{t_{\,0}}^{t_{\,1}} L({\vec q},{\dot{\vec q}};t)\,dt.
\end{equation}
 {Here the action is signed by letter $J$ instead of generally adopted $S$.
 The letter $S$ will be reserved further as a symbol for entropy, like the Boltzmann entropy.
 In turn, $J$ denotes a cost (value function, the same action~\cite{WerbosDolmatova2000}) in Bellman-Hamilton-Jacobi equation.
 It is applied in theoretical problems of robotics,
 Artificial Intelligence~\cite{FerrariStenger2004},~\cite{LaValle2006}, as well as in macroeconomics~\cite{Sargent1987}.},

 In the above equation $L({\vec q},{\dot{\vec q}};t)$ is Lagrangian describing the system "robot-maze",
 ${\vec q}$ and ${\dot{\vec q}}$ are coordinate and velocity of the robot.
 Main formulas of the Lagrangian mechanics~\cite{Lanczos1970} resulting from
 the principle of least action are collected in Table~1, page~\pageref{Table1}.

 In case of the quantum automaton, its coordinate
 ${\vec q}$ and velocity ${\dot{\vec q}}$ cannot both be accurately measured
 simultaneously.
 In addition, such a phenomenon as tunnelling across a potential barrier
 can be described only within the framework of quantum mechanics.
 It means, optimal paths from point A to B
 are quantum trajectories - {\it Bohmian trajectories}~\cite{Wyatt2005}.
 They stem from solution of modified (quantum)
 HJ equation. More strictly, they result from a right decomposition of
 the Schr{\"o}dinger equation~\cite{Bohm:1952a},~\cite{Bohm:1952b},~\cite{BohmHiley1993}.
 At present, the Bohmian interpretation of QM has a high interest from scientific
 community~\cite{Bittner2000},~\cite{Bittner2003},~\cite{ChouWyatt2007},~\cite{CoffeyEtAll2008},~\cite{Poirier2008c},
 ~\cite{Poirier2008a},~\cite{Poirier2008b},~\cite{PoirierParlant2008},
 ~\cite{TrahanPoirier2008b},~\cite{TrahanPoirier2008c},~\cite{WyattBittner2003}.

\vspace{4pt}
\parbox{16.2cm}{
{\hspace{32pt} {\label{Table1} Table 1}, {\Large \it The Legendre's dual transformations} :}
\vspace{-4pt}
\begin{center}
 \begin{tabular}{|l|l|}
   \hline
 %  & \\
   ~ {\bf \underline{Variables :}} &
   ~ {\bf \underline{Variables :}} \\
 %  & \\
   \hfil Coordinate:~ ${\vec q}=\{q_{\,1},q_{\,2},\cdots,q_{\,N}\} $ &
   \hfil Coordinate:~ ${\vec q}=\{q_{\,1},q_{\,2},\cdots,q_{\,N}\} $ \\
   & \\
    \hfil Momentum:~ ${\vec p}=\{p_{\,1},p_{\,2},\cdots,p_{\,N}\}$ &
    \hfil Velocity:~~~~~ ${\dot {\vec q}}=\{{\dot q}_{\,1},{\dot q}_{\,2},\cdots,{\dot q}_{\,N}\}$  \\
   & \\
   \hline
   \hline
  {\bf \underline{ Hamiltonian function :}} &
  {\bf \underline{ Lagrangian function :}} \\
   & \\
   \hfil ~~$H({\vec {q}},{\vec {p}};t) = \sum\limits_{n=1}^{N} {p}_{n}{\dot {q}}_{n} - L({\vec {q}},{\dot {\vec {q}}};t)$~~ &
   \hfil ~~$L({\vec {q}},{\dot {\vec {q}}};t) = \sum\limits_{n=1}^{N} {p}_{n}{\dot {q}}_{n} - H({\vec {q}},{\vec {p}};t)$~~ \\
   & \\
   \hfil ${\displaystyle{{\partial\,H}\over{\partial\,p_{\,n}}}}= {\dot {q}}_{n}$ &
   \hfil ${\displaystyle{{\partial\,L}\over{\partial\,{\dot {q}}_{\,n}}}}= p_{n}$ \\
   & \\
   \hfil ${\displaystyle{{\partial\,H}\over{\partial\,q_{\,n}}}}= -{\dot {p}}_{n}$ &
   \hfil ${\displaystyle{{\partial\,L}\over{\partial\,q_{\,n}}}}=  {\dot {p}}_{n}$ \\
   & \\
   \hline
 \end{tabular}
\end{center}
}

 The Schr{\"o}dinger equation discovered by Erwin Schr{\"o}dinger in 1926 year~\cite{Schr1926} represents a "visiting card" of quantum mechanics.
 This equation deals with wave function that is a complex function given in Hilbert space.
 David Bohm shown straightforward~\cite{Bohm:1952a},~\cite{Bohm:1952b},~\cite{BohmHiley1993},~\cite{Hiley2002}
 that the Schr{\"o}dinger equation can be decomposed onto two equations, both for real functions.
 The first equation is a modified Hamilton-Jacobi equation (HJ equation),
 and the second equation is the continuity equation.
 Quantum trajectories to be submitted to the modified HJ equation demonstrate themselves a hydrodynamical picture of the quantum mechanics~\cite{Wyatt2005}.
 This approach is very intuitive due to classical-like understanding of the underlying
 dynamics~\cite{Poirier2008c},~\cite{Poirier2008a},~\cite{Poirier2008b},~\cite{PoirierParlant2008},
 Importantly, that such approach has a pedagogical insight into entirely quantum
 mechanical effects such as tunneling and interference.

 Plural noun "paths" mentioned above at discussion of the optimal paths tracing from A to B means that there can be a set of trajectories submitting to the principle of least action~\cite{Wyatt2005},
 as for example in interferometer. The trajectory can bifurcate
 ({\it either} transmitted path {\it or} reflected path), in contrast to the Feynman's path integral paradigm~\cite{Grosche1993},~\cite{MacKenzie2000}, where a trajectory can split
 ({\it and} transmitted path {\it and} reflected path). If Bohmian trajectories are those submitted to the principle of least action, the Feynman's trajectories are all possible trajectories tracing from A to B (virtual trajectories). However, only optimal trajectories (Bohmian trajectories) survive. The others cancel each other due to interference effect.
 Short histories include created and annihilated virtual particles~\cite{Feynman1948},~\cite{FeynmanHibbs1965}.

 The Feynman's path integral formalism is
 other QM approach having a promising perspective.
 It goes back to Dirac's observation that the action plays a central role in classical
 mechanics (he considered the Lagrangian formulation of classical mechanics to be more
 fundamental than the Hamiltonian one~\cite{MacKenzie2000}).
 Initially, Dirac in papers~\cite{Dirac1933} and~\cite{Dirac1945} (1933 and 1945 years) attracted attention to a term
 $\exp(iS/{\hbar})$ corresponding to the propagator~\cite{MacKenzie2000} ($S$ is the classical action).
 The Dirac's observation provided much of the initial impetus for Feynman's work,
 making quite explicit the role of $\exp({\rm i}Ldt/\hbar)$ as a transition amplitude between states
 separated by an infinitesimal time $dt$, and its connection to the classical principle of least action.
 Feynman in 1948~\cite{Feynman1948}
 developed this idea, concerning other paths than only the classical one~\cite{FeynmanHibbs1965}.

 The path integral formalism is more intuitive and more powerful way of viewing quantum mechanics helping us
 to insight into subtle details of it.
 However, its "Achilles heel" is pathology of "infinite measure", "infinite sums of phases"
 with unit absolute values, etc~\cite{Grosche1993}.
 In particular, many of standard real-valued, time independent potentials which are used in modeling quantum systems are singular (for example, the attractive Coulomb potential) and do not fit with the theory~\cite{JohnsonLapidus2002}.
 To avoid these obstacles, DeWitt~\cite{DeWitt1957}, for example, determined a quantum corrector
 $\Delta V_{DeW}=\hbar^{\,2}/6m\cdot R$, $R$ is scalar curvature~\cite{Grosche1993}.
 It is indispensable in order to derive the  Schr{\"o}dinger equation from the time evolution integral.
% Analogous correction $\sim \hbar^{\,2}/2m$ can be achieved
 By taking into account the Bohmian quantum potential~\cite{TrahanPoirier2008b}
 we can find a corrector $\sim \hbar^{\,2}/2m$ for the path integral.
 It stems directly from the Schr{\"o}dinger equation~\cite{Bohm:1952a},~\cite{Bohm:1952b},~\cite{BohmHiley1993},~\cite{Hiley2002}.

 In this light the Bohmian amusing exercise acquires a new perception. It permits more forward steps.
 For this reason, Sect.~\ref{sec:level2} outlines emergence of the modified HJ equation and
 the continuity equation from the Schr{\"o}dinger equation.
 We repeat decomposition of the Schr{\"o}dinger equation in detail in order to get a robust base for the following computations.
 In Sect.~\ref{sec:level3} the Bohmian quantum potential is transformed to two quantum correctors.
 The first corrects the kinetic energy term and the second corrects the potential energy term.
 In Sect.~\ref{sec:level4} the modified HJ equation and the continuity equation combine together. The combined equation is complex one.
 Such a complexification generalizes the Lagrangian mechanics due to expansion of coordinates and momenta to an imaginary sector.
 Sect.~\ref{sec:level5} introduces the path integral loaded by this complexified Lagrangian because of expansion to the imaginary sector.
 Two-slit interference in light of the Bohmian trajectories is considered as an example.
 Sect.~\ref{sec:level6} gives concluding remarks and mentions the Everett's "many-worlds" theory as touching the complexified Lagrangian mechanics.

% Section2
\section{\label{sec:level2}Bohmian decomposition.}

 We begin with the following form of the The Schr{\"o}dinger equation
\begin{equation}\label{qp=2}
    {\rm i}\hbar\,{{\partial\, |\Psi({\vec {q}},{\vec {p}},t)}\rangle\over{\partial\, t}} =
    -{{\hbar}\over{2m}}\sum_{n=1}^{N} \, p_{\,n}^{\;2}\, |\Psi({\vec {q}},{\vec {p}},t)\rangle
    + U({\vec {q}})\,|\Psi({\vec {q}},{\vec {p}},t)\rangle.
\end{equation}
 Here $\hbar=h/2\pi\approx1.05457\times10^{-34}[{\rm J\cdot s}]$ is reduced Planck constant, also known as Dirac's constant,
 ${\vec {q}}=\{q_{\,1},q_{\,2},\cdots,q_{\,N}\}$ is spatial coordinate of a particle,
\begin{equation}\label{qp=3}
    {\vec  {p}} = \Biggl\{
    {{\partial~~}\over{\partial q_{\,1}}},
    {{\partial~~}\over{\partial q_{\,2}}}, \cdots ,
    {{\partial~~}\over{\partial q_{\,N}}} \Biggr\}
\end{equation}
 is momentum operator, $m$ is the particle mass, $U({\vec {q}})$ is a potential energy term. And the wave function
 $|\Psi({\vec {q}},{\vec {p}},t)\rangle$
 is a complex function of the spatial coordinate ${\vec {q}}$, momentum ${\vec {p}}$, and time $t$.
 Real function is the probability density defined by
\begin{equation}\label{qp=4}
    \rho({\vec {q}},{\vec {p}},t) =  R({\vec {q}},{\vec {p}},t)^{\,2} =
  |\Psi({\vec {q}},{\vec {p}},t)|^{\,2}= \langle\Psi({\vec {q}},{\vec {p}},t)|\Psi({\vec {q}},{\vec {p}},t)\rangle.
\end{equation}

 Without loss of generality, we express the wave function $|\Psi({\vec {q}},{\vec {p}},t)\rangle$ in terms of a real probability
 density $\rho({\vec {q}},{\vec {p}},t)$
 and a phase that depends on the real variable $J({\vec {q}},{\vec {p}},t)$
 as follows
\begin{equation}\label{qp=5}
    |\Psi({\vec {q}},{\vec {p}},t)\rangle = \sqrt{\rho({\vec {q}},{\vec {p}},t)}
    \exp\Bigl\{\,{\rm i}J({\vec {q}},{\vec {p}},t)/\hbar\,\Bigr\}=
    R({\vec {q}},{\vec {p}},t)
    \exp\Bigl\{\,{\rm i}J({\vec {q}},{\vec {p}},t)/\hbar\,\Bigr\}.
\end{equation}
 By substituting the wave function into the Schr{\"o}dinger equation~(\ref{qp=2}) we get
\begin{eqnarray}
\label{qp=6}
 &&
  \underbrace{-{{\partial J}\over{\partial\,t}}\cdot|\Psi\rangle}_{\rm (a)}
  \underbrace{+{\rm i}\hbar\,{{1}\over{2\rho}}{{\partial\,\rho}\over{\partial\,t}}\cdot|\Psi\rangle}_{\rm (b)} =
  \underbrace{{{1}\over{2m}}(\nabla J)^{2}\cdot|\Psi\rangle + U({\vec {q}})\cdot|\Psi\rangle}_{\rm (a)}
  \\
 &&
 \nonumber
  \underbrace{-{{{\rm i}\hbar}\over{2m}}\nabla^{\,2}J\cdot|\Psi\rangle
  -{{{\rm i}\hbar}\over{2m}}\Biggl({{1}\over{\rho}}\,\nabla\rho\Biggr)(\nabla J)|\Psi\rangle}_{\rm (b)}
  \underbrace{-{{\hbar^{2}}\over{2m}}\Biggl({{1}\over{2\rho}}\nabla^{\,2}\rho \Biggr)|\Psi\rangle
  +{{\hbar^{2}}\over{2m}}\Biggl({{1}\over{2\rho}}\nabla \rho \Biggr)^{2}\,|\Psi\rangle}_{\rm (c)}.
\end{eqnarray}
 Operators of gradient $\nabla$ and laplacian $\nabla^{\,2}$ read
\begin{equation}\label{qp=7}
    \nabla = \Biggr\{
    {{\partial~~}\over{\partial\,q_{\,1}}}{ i_{1}}+
    {{\partial~~}\over{\partial\,q_{\,2}}}{ i_{2}}+ \cdots +
    {{\partial~~}\over{\partial\,q_{\,N}}}{ i_{N}}
    \Biggl\}, \hskip16pt
    \nabla^{\,2} = \Biggr\{
    {{\partial^{\,2}~~}\over{\partial\,q_{\,1}^{\,2}}}+
    {{\partial^{\,2}~~}\over{\partial\,q_{\,2}^{\,2}}}+ \cdots +
    {{\partial^{\,2}~~}\over{\partial\,q_{\,N}^{\,2}}}
    \Biggl\}.
\end{equation}
 A set $\{{ i_{1}},{ i_{2}},\cdots,{ i_{N}}\}$ represents orthogonal basis of $N$-dimensional state space $S^{\,N}$.

 Collecting together real terms (a) and (c), and singly imaginary terms (b) in Eq.~(\ref{qp=6}) we obtain two coupled equations for real
 functions $J({\vec{q}},{\vec {p}},t)$ and $\rho({\vec{q}},{\vec {p}},t)$
\begin{eqnarray}
\label{qp=8}
&{\rm (a)+(c):}&  -{{\partial\,J}\over{\partial\,t}} \;=\; {{1}\over{2m}}(\nabla J)^{\,2} + U({\vec{q}}) + Q({\vec{q}},t),
  \\
 &{\rm (b):}&  -{{\partial\,\rho}\over{\partial\,t}} \;=\; \nabla \Biggl(
  \rho\;{{\nabla J}\over{m}}\Biggr).
\label{qp=9}
\end{eqnarray}
 Quantum potential $Q({\vec{q}},{\vec {p}},t)$ in~(\ref{qp=8}) measures a curvature induced by internal stress~\cite{WyattBittner2003}:
\begin{equation}\label{qp=10}
    Q  =
    -{~{\hbar^{\,2}}\over{2m}} \Biggl[\;{{\nabla^{\,2}\rho}\over{2\rho}} -
    \Biggl({{\nabla\rho}\over{2\rho}}\Biggr)^{\,2\,}\Biggr]
   = -{~{\hbar^{\,2}}\over{2m}}{{\nabla^{\,2}R}\over{R}}
\end{equation}
% measures the curvature induced internal stress~\cite{WyattBittner2003}.
 The above equations,~(\ref{qp=8}) and~(\ref{qp=9}), are seen to be
 the coupled pair of nonlinear partial differential equations~\cite{Poirier2008c},~\cite{Poirier2008a},~\cite{Poirier2008b},~\cite{TrahanPoirier2008b}.
 The first of the two equations, Eq.~(\ref{qp=8}), is the Hamilton-Jacobi equation modified by the quantum potential $Q({\vec{q}},{\vec {p}},t)$.
 The second equation, Eq.~(\ref{qp=9}), is the continuity equation.

 Momentum of the particle is
\begin{equation}\label{qp=11}
    {\vec {p}} = m {\vec {v}} = \nabla\,J,
\end{equation}
 where ${\vec {v}}$ is its velocity. And
\begin{equation}\label{qp=12}
    {{1}\over{2m}}(\nabla\,J)^{2} =
    {{1}\over{2m}}\,{\vec {p}}^{\,2}
\end{equation}
 is the kinetic energy of the particle.
 The particle's energy is $E=-\partial\,J/\partial\,t$.
 Equation (\ref{qp=8}) states that total energy is the sum of the kinetic energy, potential energy, and the quantum potential~\cite{Hiley2002}.
 Equation (\ref{qp=9}), in turn, is interpreted as simply the continuity equation for probability density $\rho({\vec q},{\vec {p}},t)$.
 It says that all individual trajectories demonstrate collective behavior like a liquid flux~\cite{Lanczos1970},~\cite{Wyatt2005}, perhaps, superconductive one.

 We shall see that the quantum potential $Q({\vec {q}},{\vec {p}},t)$ corrects both the kinetic energy term and the potential energy term.
 Therefore, further the quantum potential corrector $Q({\vec {q}},{\vec {p}},t)$ will be called simply as the {\it quantum corrector}.

% Section3
\section{\label{sec:level3}The quantum corrector as an information channel.}

 According to the observation
 \begin{equation}\label{qp=13}
    \rho^{-1}\cdot\nabla \rho =\nabla \ln(\rho)
 \end{equation}
 we can prepare the quantum corrector by the following way~\cite{Sbitnev2008a}
\begin{eqnarray}
\nonumber
  Q({\vec {q}},t) &=& {{\hbar^{2}}\over{2m}}\Biggl[\biggl(\,{{1}\over{2\rho}}\,\nabla \rho \biggr)^{2}
   -{{1}\over{2}}\biggr({{1}\over{\rho}}\nabla \biggr(\,\rho\cdot{{1}\over{\rho}}\,\nabla \rho \biggl)\biggl)\Biggl] \\
\nonumber
    &=& {{\hbar^{2}}\over{2m}}\Biggl[\biggl(\,{{1}\over{2}}\nabla \ln(\rho) \biggr)^{2}
    -{{1}\over{2}}\biggr(\,{{1}\over{\rho}}\,\nabla \biggr(\rho\, \nabla \ln(\rho) \biggl)\biggl)\Biggl] \\
\nonumber
    &=& {{\hbar^{2}}\over{2m}}\Biggl[\biggl(\,{{1}\over{2}}\nabla \ln(\rho) \biggr)^{2}
     -{{1}\over{2}}\biggr(\,{{1}\over{\rho}}\,\nabla \rho\cdot\nabla \ln(\rho) + {{\rho}\over{\rho}}\cdot\nabla^{\,2} \ln(\rho)\biggl) \Biggl] \\
%\nonumber
     &=&
     \textcolor{magenta}{
     -{{\hbar^{2}}\over{2m}}\Biggl(\,{{1}\over{2}}\,\nabla \ln(\rho) \Biggr)^{2}
        -{{\hbar^{2}}\over{2m}}\Biggl(\,{{1}\over{2}}\,\nabla^{\,2} \ln(\rho) \Biggr) }
\label{qp=14}
\end{eqnarray}
 Define a logarithmic function
\begin{equation}\label{qp=15}
    S_{Q}({\vec {q}},{\vec {p}},t)=-{{1}\over{2}}\,\ln(\rho({\vec {q}},{\vec {p}},t))
    =-\ln\biggr(\sqrt{\rho({\vec {q}},{\vec {p}},t)}\biggl)
\end{equation}
 to be called further {\it quantum entropy}.
 It is like to the Boltzmann entropy.
 It characterizes degree of order and chaos of some
 entity (vacuum, holomovement~\footnote[1]{~See
 \urlprefix\url{http://en.wikipedia.org/wiki/Holomovement} ,
% $\Rightarrow$ "Holomovement"
 see also~\cite{Hiley2002}.})
 supporting $\rho({\vec {q}},{\vec {p}},t)$.
 Observe that, vacuum is a storage of virtual trajectories supplying
 optimal ones for particle movement~\cite{FeynmanHibbs1965}.

 Substituting $S_{Q}({\vec {q}},{\vec {p}},t)$ into Eq.~(\ref{qp=14})
 we obtain the quantum corrector expressed in terms of
 this logarithmic function
 {(in~\cite{Bittner2003},~\cite{Wyatt2005},~\cite{WyattBittner2003}
  the term $-S_{Q}$ (negative $S_{Q}$) is named $C$-amplitude)}
\begin{equation}\label{qp=16}
    Q({\vec {q}},{\vec {p}},t) =
    \underbrace{\textcolor{magenta}{
    -{{\hbar^{2}}\over{2m}}\,(\nabla S_{Q})^2}}_{\rm (a)}
    \underbrace{\textcolor{magenta}{
    +{{\hbar^{2}}\over{2m}}\,\nabla^{\,2}S_{Q}}}_{\rm (b)}.
\end{equation}
 Here the term enveloped by brace (a) is viewed as the quantum corrector of the kinetic energy term.
 And the term enveloped by brace (b) corrects the potential energy term. Later on we will analyze this correction. But now let us substitute
 this quantum corrector to Eq.~(\ref{qp=8})
\begin{equation}\label{qp=17}
    -{{\partial J}\over{\partial\,t}} =
    \underbrace{
     {{1}\over{2m}}(\nabla J)^{2}
     \textcolor{magenta}{
     -{{\hbar^{2}}\over{2m}}(\nabla S_{Q})^{2}}}_{\rm (a)}
    \underbrace{
    + U({\vec {q}})
    \textcolor{magenta}{
    + {{\hbar^{2}}\over{2m}}\nabla^{\,2}S_{Q}}}_{\rm (b)}.
\end{equation}
 Terms enveloped by brace (a) relate to the kinetic energy term, and those enveloped by brace (b) relate to the potential energy term.
 Terms colored in magenta are the quantum correctors.
 Observe that, by handling with the entropy $S_{Q}$ instead of $\rho$
 we get a useful way to transform a non-linear model to a linear
  one~\cite{Bittner2000}.
 Substituting also $S_{Q}$ in the continuity equation~(\ref{qp=9})
 instead of $\rho$ we obtain the entropy balance equation
\begin{equation}\label{qp=18}
   {{\partial S_{Q}}\over{\partial\,t}} =
   -({\vec {v}}\cdot \nabla S_{Q})
   +
     {{1}\over{2}}(\nabla\,{\vec {v}}).
\end{equation}
 Here ${\vec {v}}=\nabla J/m$ is the particle speed. A single term $(\nabla\,{\vec {v}})$ describes a rate of the entropy flow due to
 spatial divergence of the speed.
 This term is nonzero in regions where the particle
 changes direction of movement.
 Observe that negative $S_{Q}$, $C$-amplitude~\cite{WyattBittner2003}~\cite{Wyatt2005}, relates to information~\cite{Brillouin2004}. So, this equation describes balance of the information flows.

% Section4
\section{\label{sec:level4}Beyond the Bohm's insight into QM core.}

 Pair of the equations - the modified HJ equation~(\ref{qp=17}) and the entropy balance equation~(\ref{qp=18}),
 describes behavior of the quantum particle exactly~\cite{TrahanPoirier2008b}. Let us now multiply Eq.~(\ref{qp=18})
 by the factor $-{\rm i}\hbar$
 and add the result to Eq.~(\ref{qp=17}). We obtain the following complex HJ equation
\begin{eqnarray}
\nonumber
    -{{\partial \Jc}\over{\partial\,t}} &=&
       \underbrace{
     {{1}\over{2m}}(\nabla J)^{2}
     + {\rm i}\hbar\,{{1}\over{m}}(\nabla J\cdot \nabla S_{Q})
     \textcolor{magenta}{-{{\hbar^{2}}\over{2m}}(\nabla S_{Q})^{2}}}_{\rm (a)}
     \\
\label{qp=19}
    &+&
     \underbrace{ U({\vec {q}})
    -  \textcolor{blue}{{\rm i}\hbar\,{{1}\over{2}}(\nabla\,{\vec {v}})}
    + \textcolor{magenta}{{{\hbar^{2}}\over{2m}}\nabla^{\,2}S_{Q}}}_{\rm (b)}
\end{eqnarray}
 Here the term %$\Jc=J+{\rm i}\hbar S_{Q}$
\begin{equation}\label{qp=20}
    \Jc=J+{\rm i}\hbar S_{Q}
\end{equation}
 is {\it complexified action}.
 Terms enveloped by brace (a) can be rewritten
 as gradient of the complexified action squared
\begin{equation}\label{qp=21}
    {{1}\over{2m}}(\nabla\Jc)^{2}=
    {{1}\over{2m}}(\nabla J)^{2}
     +{\rm i}\hbar\,{{1}\over{m}}(\nabla J\cdot \nabla S_{Q})
     -{{\hbar^{2}}\over{2m}}(\nabla S)^{2}
\end{equation}
 As for the terms enveloped by brace (b) they could stem from expansion into the Taylor's series
 of the potential energy extended previously to a complex space, see, for example, like complex extension in~\cite{Poirier2008c}. In our case,
 the potential function is extended in the complex space possessing by a small broadening into imaginary sector.
 In this respect, let us now expand into the Taylor's series the potential function having a complex argument~\footnote[2]{~Taylor series
 can also be defined for functions of a complex variable. It follows from applying the Cauchy integral formula:
 \urlprefix\url{http://mathworld.wolfram.com/TaylorSeries.html} .}
\begin{equation}\label{qp=22}
    U({\vec {q}}+{\rm i}{\vec \epsilon}\,) \approx  U({\vec {q}})
    + \textcolor{blue}{ {\rm i}{\vec \epsilon}\; \nabla U({\vec {q}})}
    \textcolor{magenta}{ - }
    \textcolor{magenta}{ {{\epsilon^{\,2}}\over{2}}\; \nabla^{\,2} U({\vec {q}})} + \cdots
\end{equation}
 Terms colored in blue and magenta relate to the same terms in Eq.~(\ref{qp=19}).
 Let us examine their.
 Here a small vector ${\vec \epsilon}$ has dimensionality of length.
 But it should contain also the Planck constant, $\hbar$, to
 reproduce the second and third terms enveloped by brace (b) in Eq.~(\ref{qp=19}).
% Let us write down this vector as follows
 A minimal representation of this vector can be as follows
\begin{equation}\label{qp=23}
    {\vec \epsilon} = {{\,\hbar}\over{2m}}\,s\,{\vec n}.
\end{equation}
 Here $m$ is the particle's mass, and $s$ is the universal constant,
 "reverse velocity"~\cite{Poluyan2005},
\begin{equation}\label{qp=24}
    s = 4\pi \varepsilon_{0}\,{{\hbar}\over{e^{\,2}}}\approx 4.57\times10^{-7}~[{\rm s/m}],
\end{equation}
 where $e\approx-1.6\times10^{-19}~[{\rm C}]$ is the elementary charge carried by a single electron and
 $\varepsilon_{0}\approx8.854\times10^{-12}~[{\rm C^{\,2}\,N^{-1}m^{-2}}]$ is the vacuum permittivity.
 Observe that, multiplication of the constant $s$ by the speed of light $c$ gives
 enigmatic universal dimensionless constant~\footnote[3]{~
 \urlprefix\url{http://en.wikipedia.org/wiki/Fine-structure_constant} :
 $\alpha=7.297352570\times10^{-3}\approx1/137.035999070$.}
 close to $137^{-1}$. Why did we choose the constant $s$ but not $c^{-1}$?
 The answer is that the speed $c$ deals with the relativistic movements, but the constant $s$ relates directly
 to quantum realm.
 As for the vector ${\vec n}$ in Eq.~(\ref{qp=23}) it we believe is a unit vector, $(\vec n,\vec n)=1$,
 i.e., any variation of ${\vec \epsilon}$ is permitted on an $N$-dimensional sphere having unit radius.
 In light of these remarks, we can rewrite the expansion~(\ref{qp=22}) in the following form
\begin{equation}\label{qp=25}
    U({\vec {q}}+{\rm i}{\vec \epsilon}\,) \approx
    U({\vec {q}})
    + {\underbrace{ \textcolor{blue}{ {\rm i}\hbar\Biggl({\vec n}\cdot\Biggl( {{s}\over{2m}} \nabla U({\vec {q}})\Biggr)\Biggr)} }_{\rm (b_1)}}
     {\underbrace{\textcolor{magenta}{ - {{\hbar^{\,2}}\over{2m}}\Biggl( {{s^{\,2}}\over{2m}} \nabla^{\,2} U({\vec {q}}) \Biggr)} }_{\rm (b_2)}} + \cdots
\end{equation}
 A therm enveloped by brace (b$_1$) contains unit vector ${\vec n}$ pointing out direction of the imaginary broadening.
 A force ${\vec F}=-\nabla U({\vec {q}})$ multiplied by $l{\vec n}$ is elementary work, performing by this force at shifting on a length $l$
 along ${\vec n}$.
 The force multiplied by the factor $s{\vec n}$ and divided into mass $m$ is a rate of velocity's variation per unit length,
 i.e., it represents divergence of the velocity, $(\nabla\,{\vec {v}})$. So, the term enveloped by brace (b$_1$) can be rewritten
 in the following form
\begin{equation}\label{qp=26}
    {\rm (b_{1}):}~~~~
   \textcolor{blue}{ {{s}\over{2m}}\Bigl({\vec n}\cdot\nabla U({\vec {q}})\Bigr) = -{{1}\over{2}}(\nabla\cdot{\vec {v}})}.
\end{equation}
 As for a term $(s^{\,2}/2m)\cdot U({\vec {q}})$ shown over brace (b$_2$) in Eq.~(\ref{qp=25}) it is dimensionless.
 Accurate to an additive dimensionless term $a{\vec q}^{\,2}+({\vec b}{\vec q})+c$ it is comparable with $S_{Q}$.
 We proclaim
\begin{equation}\label{qp=27}
    {\rm (b_2):}~~~~
  \textcolor{magenta}{  -\Biggl({{s^{\,2}}\over{2m}}\nabla^{\,2} U({\vec {q}}) \Biggr) = \nabla^{\,2} S_{Q}}.
\end{equation}

 Further we shall consider complexified momentum
\begin{equation}\label{qp=28}
    {\vec {\Pc}}=m{\dot {\vec {\Qc}}} = \nabla \Jc
    =\nabla J+{\rm i}\,\hbar\,\nabla S_{Q}
\end{equation}
 and complexified coordinate
\begin{equation}\label{qp=29}
    {\vec {\Qc}}={\vec {q}}+{\rm i}\,{\vec \epsilon}
\end{equation}
 as extended representations of the real vectors ${\vec {p}}$ and ${\vec {q}}$.
 The complexified momentum~${\vec {\Pc}}$ differs from the momentum ${\vec p}$ by additional imaginary term $\hbar\,\nabla S_{Q}$.
 And the complexified coordinate ${\vec {\Qc}}$ differs from the real coordinate ${\vec {q}}$ by the small imaginary vector~(\ref{qp=23}).
 Now we can rewrite Eq.~(\ref{qp=19}) as compexified the Hamilton-Jacobi equation:

\vspace{8pt}
\noindent
\fbox{
\parbox{17.6cm}{
\begin{equation}\label{qp=30}
       -{{\partial \Jc}\over{\partial\,t}} =
       {{1}\over{2m}}(\nabla\Jc)^{2}
       + U({\vec {\Qc}})
       = \Hc({\vec {\Qc}},{\vec {\Pc}};t) .
\end{equation}
}}\vspace{8pt}

\noindent
 From the right side complexified Hamiltonian $\Hc({\vec {\Qc}},{\vec {\Pc}};t)$ is placed.

 Observe that the total derivative of the complex action is as follows
\begin{equation}\label{qp=31}
    {{d\Jc}\over{d\,t}} \;=\; {{\partial\Jc}\over{\partial\,t}} +
    \sum\limits_{n=1}^{N} {{\partial\Jc}\over{\partial{\Qc}_{n}}}{{d{\Qc}_{n}}\over{d\,t}} \;=\;
    {{\partial\Jc}\over{\partial\,t}} + \sum\limits_{n=1}^{N} {\Pc}_{n}{\dot {\Qc}_{n}}
\end{equation}
 where complex derivative reads (see Ch.2 in~\cite{Titchmarsh1976}, for example)
\begin{equation}\label{qp=32}
  {{\partial\Jc}\over{\partial{\Qc}_{n}}} \;=\; {{\partial J}\over{\partial q_{n}}}
  + {\rm i}\,\hbar\, {{\partial S_{Q}}\over{\partial q_{n}}} \;=\; {\Pc}_{n}.
\end{equation}
 Combining Eq.~(\ref{qp=31}) with~(\ref{qp=30}) we obtain the the Legendre's dual transformation \cite{Lanczos1970} binding the Hamiltonian $\Hc$ and the Lagrangian $\Lc$

\vspace{8pt}
\noindent
\fbox{
\parbox{17.6cm}{
\begin{equation}\label{qp=33}
    {{d\Jc}\over{d\,t}} \;=\; -\Hc({\vec {\Qc}},{\vec {\Pc}};t)
    +  \sum_{n=1}^{N} {\Pc}_{n}{\dot {\Qc}}_{n} \;=\;
    \Lc({\vec {\Qc}},{\dot {\vec {\Qc}}};t).
\end{equation}
}}\vspace{8pt}

 We summarize this section by collecting in Table~2 the Legendre's dual transformations
 of the above complexified Hamiltonian and Lagrangian functions:

\vspace{4pt}
\parbox{16.2cm}{
{\hspace{32pt} {\label{Table2} Table 2}, {\Large \it The Legendre's dual transformations} :}
\vspace{-4pt}
\begin{center}
 \begin{tabular}{|l|l|}
   \hline
 %  & \\
   ~ {\bf \underline{Variables :}} &
   ~ {\bf \underline{Variables :}} \\
 %  & \\
   \hfil Coordinate:~ ${\vec \Qc}={\vec q}+{\rm i}{\displaystyle{{\,\hbar}\over{2m}}}\,s\,{\vec n} $ &
   \hfil Coordinate:~ ${\vec \Qc}={\vec q}+{\rm i}{\displaystyle{{\,\hbar}\over{2m}}}\,s\,{\vec n} $ \\
   & \\
    \hfil Momentum:~ ${\vec \Pc}={\vec p}+{\rm i}\hbar\nabla S_{Q}$~~ &
    \hfil Velocity:~~~~~ ${\dot {\vec \Qc}}={\dot {\vec q}}+{\rm i}{\displaystyle{{\,\hbar}\over{2m}}}\,s\,{\dot {\vec n}}$  \\
   & \\
   \hline
   \hline
  {\bf \underline{ Hamiltonian function :}} &
  {\bf \underline{ Lagrangian function :}} \\
   & \\
   \hfil ~~$\Hc({\vec {\Qc}},{\vec {\Pc}};t) = \sum\limits_{n=1}^{N} {\Pc}_{n}{\dot {\Qc}}_{n} - \Lc({\vec {\Qc}},{\dot {\vec {\Qc}}};t)$~~ &
   \hfil ~~$\Lc({\vec {\Qc}},{\dot {\vec {\Qc}}};t) = \sum\limits_{n=1}^{N} {\Pc}_{n}{\dot {\Qc}}_{n} - \Hc({\vec {\Qc}},{\vec {\Pc}};t)$~~ \\
   & \\
   \hfil ${\displaystyle{{\partial\,\Hc}\over{\partial\,\Pc_{\,n}}}}= {\dot {\Qc}}_{n}$ &
   \hfil ${\displaystyle{{\partial\,\Lc}\over{\partial\,{\dot {\Qc}}_{\,n}}}}= \Pc_{n}$ \\
   & \\
   \hfil ${\displaystyle{{\partial\,\Hc}\over{\partial\,\Qc_{\,n}}}}= -{\dot {\Pc}}_{n}$ &
   \hfil ${\displaystyle{{\partial\,\Lc}\over{\partial\,\Qc_{\,n}}}}=  {\dot {\Pc}}_{n}$ \\
   & \\
   \hline
 \end{tabular}
\end{center}
}

\noindent
 The Lagrangian equations of motions and the Legendre's transformations~\cite{Lanczos1970} are  invariant
 under the above imaginary extension of the real momenta, $p_{\,n}$, and the real velocities, $v_{\,n}$, ($n=1,2,\cdots,N$).
 It should be noted, that the Hamiltonian function is quadratic in the momenta, $\Pc_{n}$,
 and the Lagrangian function is quadratic in the velocities, ${\dot {\Qc}}_{n}$.
 A conservation law in this case unifies conservation of energy represented by real part, ${\rm Re}[\Hc({\vec {\Qc}},{\vec {\Pc}};t)]$,
 and the entropy balance~(\ref{qp=18}) represented by imaginary part, ${\rm Im}[\Hc({\vec {\Qc}},{\vec {\Pc}};t)]$.

 Turning back to Eq.~(\ref{qp=30}) one can write an action solution
\begin{equation}\label{qp=34}
    \Jc = -\int\limits_{t_{0}}^{t_{}}\, \Hc( {\vec {\Qc}},{\vec {\Pc}};\tau )d\tau + C_{1}.
\end{equation}
 On the other hand the same solution can be obtained by integrating Eq.~(\ref{qp=33})
\begin{equation}\label{qp=35}
    \Jc = \int\limits_{t_{0}}^{t_{}}\, \Lc( {\vec {\Qc}},{\dot{\vec {\Qc}}};\tau )d\tau + C_{2}.
\end{equation}
 Here $C_{1}$ and $C_{2}$ are integration constants relating to different particular integrals.
 One can see the both integration constants satisfy the following condition
\begin{equation}\label{qp=36}
    C_{1}-C_{2}= \int\limits_{t_{0}}^{t_{}}\;
    \sum\limits_{n=1}^{N} {\Pc}_{n}\,{\dot {\Qc}}_{n}\,dt
    = \int\limits_{L}\;
    \sum\limits_{n=1}^{N} {\Pc}_{n}\,{d{\Qc}}_{n}.
\end{equation}
 Here $L$ is a curve beginning at $t_{0}$ and terminating at $t$. It is given in the state space $S^{\,N}$.
 Observe that the curvilinear integral along a closed curve
\begin{equation}\label{qp=37}
    {\mit\Gamma} = \oint \sum\limits_{n=1}^{N} {\Pc}_{n}\,{d{\Qc}}_{n}
\end{equation}
 is an {\it invariant of the motion}: ${\mit\Gamma}={\rm const}$~\cite{Lanczos1970}. From here it follows, in particular, if a circulation around every closed curve is zero at $t=0$, then the same property holds permanently. This means that a "quantum fluid" which is initially free from vortices remains so permanently, i.e., vortices cannot be created or destroyed.

 As follows from Eq.~(\ref{qp=28}) we have
\begin{equation}\label{qp=38}
    {\dot {\vec {\Qc}}} = {\dot {\vec {q}}} + {\rm i} {\dot{\vec \epsilon}} \;=\;
    {{1}\over{m}} {\vec {\Pc}} \;=\;
    {{1}\over{m}}\,\nabla J + {\rm i}{{\hbar}\over{m}}\,\nabla S_{Q}.
\end{equation}
 By comparing imaginary parts in this equation we find
\begin{equation}\label{qp=39}
    {\dot{\vec \epsilon}} = {{\,\hbar}\over{2m}}\,s\,{\dot{\vec n}}
    = {{\hbar}\over{m}}\,\nabla S_{Q} ~~~~\Rightarrow ~~~~
        {{s}\over{2}}\,{\dot{\vec n}} = \nabla S_{Q}.
\end{equation}
 The unit vector ${\vec n}$ remains always on the unit sphere. Therefore, independently of its variation,
 ${\dot{\vec n}}{\delta t} \approx {\vec n}(t+\delta t)-{\vec n}(\,t\,)$,
 its tip undergoes rotations on the unit sphere
 under the quantum entropy's variations, $\nabla S_{Q}$.
 The quantum entropy $S_{Q}$ undergoes variations within regions where
 the potential $U({\vec q})$ varies.

 Under these variations, tip of the small vector ${\dot{\vec \epsilon}}$ makes rotating movements on the sphere of a radius $r=s\hbar/2m$.
 This radius is about $2.6\times10^{-11}\;{\rm [m]}$ for electron and it is about $1.4\times10^{-14}\;{\rm [m]}$ in case of proton and neutron.
 In particular, given rest mass $m=1\,[{\rm mg}]=10^{-6}\,[{\rm kg}]$ the radius will be about
 the Planck length, $1.6\times10^{-35}\;{\rm [m]}$.
 At large masses the radius collapses and we return to semiclassical realm.

 \textcolor[rgb]{0.00,0.00,1.00}
 {A direction expanded beyond the real coordinate space $S^{\,N}$ 
 - the imaginary direction - is essential character of the quantum realm}.

% Section5
\section{\label{sec:level5}The path integral on a complexified state space.}

 Solution of the Schr{\"o}dinger equation is the following exponent
\begin{equation}\label{qp=40}
    |\,\Psi({\vec {\Qc}},{\vec {\Pc}},t)\,\rangle
  = \exp\Biggl\{\;{{\rm i}\over{\hbar}}\,\Jc\;\Biggr\}
  = \exp\Biggl\{\,{{\rm i}\over{\hbar}}\,J - S_{Q}\,\Biggr\}.
\end{equation}
 By substituting the action integral~(\ref{qp=34}) into this exponent  we get
\begin{equation}\label{qp=41}
 |\,\Psi({\vec {\Qc}},{\vec {\Pc}},t)\,\rangle =
 {{1}\over{Z_{1}}}
    \exp\Biggl\{-{{\rm i}\over{\hbar}}
    \int\limits_{t_{0}}^{t_{}} \Hc({\vec {\Qc}},{\vec {\Pc}};\tau)\,d\tau\,
    \Biggr\}.
\end{equation}
 Here $Z_{1}=\exp(-{\rm i}/\hbar\cdot C_{1})$.
 Probability density calculated is
\begin{equation}\label{qp=42}
    \langle\,\Psi({\vec {\Qc}},{\vec {\Pc}},t)\,|\,\Psi({\vec {\Qc}},{\vec {\Pc}},t)\,\rangle = \exp\Bigl\{\, - 2S_{Q}\,\Bigr\}
    = \exp\Bigl\{\, \ln(\rho({\vec q},{\vec {p}},t))\,\Bigr\} = \rho({\vec q},{\vec {p}},t).
\end{equation}

 The most fundamental quantity in the mathematical analysis of mechanical problems is the
 {\it Lagrangian function}~\cite{Arnold1978},~\cite{Lanczos1970}.
 It is defined as the excess of kinetic energy over potential energy.
 With this definition in mind we can enunciate d'Alembert's principle, $\delta\Jc=0$.
 This is  "Hamilton's principle"~\cite{Lanczos1970}.
 It states that the motion of an arbitrary mechanical system
 occurs in such a way that
     \textcolor{blue}{
     {\it definite integral}~(\ref{qp=35}) {\it becomes stationary for arbitrary possible variations of the configuration of the system,
   provided the initial and final configurations of the system are prescribed}}.
 This principle can be reformulated with respect to a form
\begin{equation}\label{qp=43}
    |\,\Psi({\vec {\Qc}},{\vec {\Pc}},t)\,\rangle \;=\;
    \exp\Biggl\{\;{{\rm i}\over{\hbar}}\,\Jc\;\Biggr\} =
    {{1}\over{Z_{2}}}\exp\Biggl\{\;{{\rm i}\over{\hbar}}\,
    \int\limits_{t_{0}}^{t_{}}\, \Lc( {\vec {\Qc}},{\dot{\vec {\Qc}}};\tau )d\tau
    \;\Biggr\}.
\end{equation}
 Here $Z_{2}=\exp(-{\rm i}/\hbar\cdot C_{2})$.
 In this case the principle states:
   \textcolor{blue}{
 {\it this exponent becomes stationary for arbitrary possible variations of the configuration of the system, provided the initial and final configurations of the system are prescribed}}. Obviously, it results from  stationarity of the integral~(\ref{qp=35}) stated above.
\begin{figure}
  \begin{picture}(400,420)(46,15)
  \centering
  \includegraphics{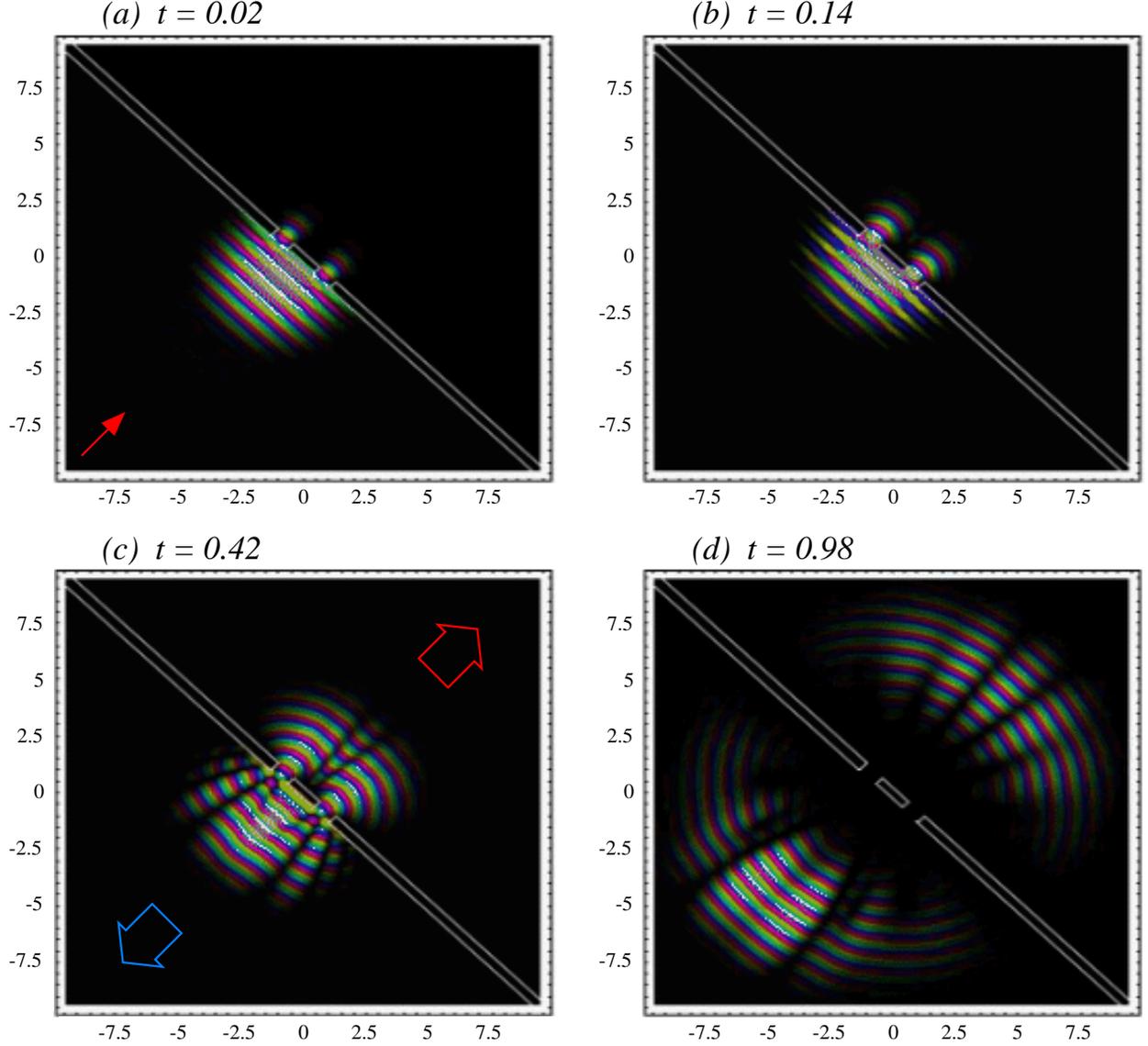}
  \end{picture}
  \caption{
    Scattering gaussian wave packet on a two-slit screen.
    Red arrow in (a) shows its initial movement directed to the screen.
    Red and blue arrows in (c) show directions of %scattering
    transmitted and reflected waves.
    Slides have been captured from a movie gif-file
    fulfilled by Max Sukharev and shown in the site
    \urlprefix\url{http://phorum.lebedev.ru/viewtopic.php?t=14}.
  }
  \label{fig=1}
\end{figure}
\begin{figure*}
  \begin{picture}(200,70)(45,15)
  \centering
  \includegraphics[scale=0.75]{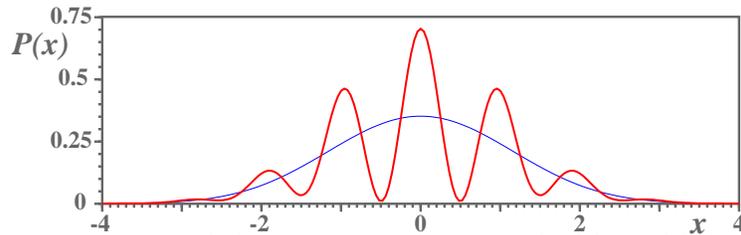}
  \end{picture}
  \caption
  {
    Red curve shows
    probability density~(\ref{qp=45}).
    Blue curve draws middle $(\rho_{\,1}(x) + \rho_{\,2}(x))/2$.
    It is a curve spanned by brace (a) in~(\ref{qp=45}) and divided by 2.
    These curves are drawn at $\sigma=1$ and $x_{\,1}=-x_{\,2}=0.5$.
  }
  \label{fig=2}
\end{figure*}

 Fundamental principle from quantum mechanics, principle of superposition, says that sum of the wave functions
 $|\,\Psi_{\,l}({\vec {\Qc}},{\vec {\Pc}},t)\,\rangle$, $l=1,2,\cdots$, represents a solution of a quantum-mechanical system as well.
 A probability density relating to superposition of two wave functions tracing two different paths
 (trajectories can bifurcate to different routes in a double-slit experiment before they are recombined at a detector)
 is
\begin{eqnarray}\label{qp=44}
  && {{1}\over{2}}\sum_{l=1}^{2} \langle \,\Psi_{\,l}({\vec {\Qc}},{\vec {\Pc}},t)\,| \cdot
     {{1}\over{2}}\sum_{k=1}^{2} |\,\Psi_{\,k}({\vec {\Qc}},{\vec {\Pc}},t)\,\rangle \\
\nonumber
  &=& {{1}\over{4}}\Bigl(\exp\Bigl\{-2S_{Q;\,1}\Bigr\} + \exp\Bigl\{-2S_{Q;\,2}\Bigr\}
  +      2\exp\Bigl\{-S_{Q;\,1}-S_{Q;\,2}\Bigr\}
    \overbrace{ \cos\bigl((J_{1}-J_{2})/\hbar \bigr) }^{\rm Interference~term}\, \Bigr).
\end{eqnarray}
 Cosine in this formula can vary between $1$ and $-1$ depending of the difference $J_{1}-J_{2}$.
 It means, that limiting values of the probability density are
$$
  {{1}\over{4}}\Bigl(\exp\{-S_{Q;\,1}\}+\exp\{-S_{Q;\,2}\}\Bigr)^{\,2}~~~~~{\rm and}~~~~~
  {{1}\over{4}}\Bigl(\exp\{-S_{Q;\,1}\}-\exp\{-S_{Q;\,2}\}\Bigr)^{\,2},
$$
 respectively.

 Figure~\ref{fig=1} shows scattering gaussian wave packet on a screen containing two slits. Interference pattern far from the screen,
 in the Fraunhofer zone, becomes apparent.
 It simulates interference of electron on two slits~\cite{Morozov2005} got at calculating the Schr{\"o}dinger equation.

 In turn, probability density relating to the interference pattern far from the slits
\begin{equation}\label{qp=45}
    P(x) = {{1}\over{2}}
 \Bigl(\,
 {\underbrace{\rho_{\,1}(x) + \rho_{\,2}(x)}_{(a)}}
 + 2\sqrt{\rho_{\,1}(x)\rho_{\,2}(x)}\cdot\cos(kx)\Bigr)
\end{equation}
 is drawn in Figure~\ref{fig=2} by red curve.
 Here $\rho_{\,1}(x)$ and $\rho_{\,2}(x)$ are described by normal distribution
\begin{equation}\label{qp=46}
   \rho_{\,l}(x)={{1}\over{\sigma\sqrt{2\pi}}}\exp\Biggl\{
   -{{(x-x_{l})^2}\over{2\sigma^{2}}}
   \Biggr\}.
\end{equation}
 The slits $l=1,2$ are $\Lambda=(x_{\,1}-x_{\,2})$ apart. And $\Lambda$ is multiple of wavelength $\lambda=2\pi/k$,
 where $k$ is the angular wave number. Curves in Fig.~\ref{fig=2} are drawn at $\sigma=1$ and $x_{\,1}=-x_{\,2}=0.5$.
 Observe that interference terms can depress different wave functions of the set
 $\{|\,\Psi_{\,l}({\vec {\Qc}},{\vec {\Pc}},t)\,\rangle$, $l=1,2,,\cdots\}$
 except for that relating to a trajectory satisfying the principle of least action.

 Feynman's brilliant revelation based on the superposition principle
 is that all arbitrary trajectories are accepted as possible histories of the evolving quantum system.
 Contributions of most paths to the integral~(\ref{qp=35}) will cancel each other,
 unless these paths are somehow "close" to the solution of $\delta \Jc = 0$, which is
 the real path of the system. In the semiclassical region the propagator will therefore be
 dominated by those paths which are in the immediate vicinity of the classical path;
 the size of this vicinity follows from the estimate $\delta \Jc \sim \hbar$. Mathematically,
 the Feynman's revelation, in cartesian coordinates (with obvious lattice discretization), reads~\cite{Grosche1996}
\begin{equation}\label{qp=47}
    G({\vec \Qc}{\,''},t;{\vec \Qc}{\,'},t_{\,0})=
 {\overbrace{\underbrace{
       \;\int\int\cdots\int\;
 }_{{\vec {\Qc}}(t_{0})={\vec {\Qc}}{\,'}}}^{ {\vec {\Qc}}(t_{})={\vec {\Qc}}{\,''}}}
       \,\Dc[{\vec \Qc}(\tau)]\,
          \exp\Biggl\{\;{{\rm i}\over{\hbar}}\,
    \int\limits_{t_{0}}^{t_{}}\, \Lc( {\vec {\Qc}},{\dot{\vec {\Qc}}};\tau )d\tau
    \;\Biggr\}.
\end{equation}
 where the path-integral symbol indicates the multiple integral~\cite{Weigel1986}
\begin{equation}\label{qp=48}
 {\overbrace{\underbrace{
       \;\int\int\cdots\int\;
 }_{{\vec {\Qc}}(t_{0})={\vec {\Qc}}{\,'}}}^{ {\vec {\Qc}}(t_{})={\vec {\Qc}}{\,''}}}
%    {\int\int\cdots\int }
    \,\Dc[{\vec \Qc}(\tau)]\;\Leftrightarrow\;
    (2\pi{\rm i}\,\hbar\,{\delta t}/m)^{-M/2}
    \int\limits_{{\vec {\Qc}}{\,'}}^{~{\vec {\Qc}}{\,''}}d{\vec \Qc}_{1}
    \int\limits_{{\vec {\Qc}}{\,'}}^{~{\vec {\Qc}}{\,''}}d{\vec \Qc}_{2}
    \cdots
    \int\limits_{{\vec {\Qc}}{\,'}}^{~{\vec {\Qc}}{\,''}}d{\vec \Qc}_{M}
\end{equation}
 in the limit ${\delta t}\rightarrow 0$, $M\rightarrow \infty$, $M{\delta t}=t-t_{0}$.
 The factor $(2\pi{\rm i}\,\hbar\,{\delta t}/m)$ contains $m$ that is the particle's mass. Thus, dimensionality of this factor is
 $[{\rm length}^{\,2}]$. And dimensionality of the normalizing factor $(2\pi{\rm i}\,\hbar\,{\delta t}/m)^{-M/2}$ is
 $[{\rm length}^{\,2}]^{-M/2}=[{\rm length}]^{-M}$.

 Fundamental principle from quantum mechanics, principle of superposition, underlies  the path integral.
 Whereas evolution of a classical object is described  by a unique trajectory satisfying the principle of least action,
 the path integral tests all possible virtual classical trajectories, among which there is a unique trajectory satisfying the least action principle. Other trajectories cancel each other by their interference.

 Feynman's path integral represented in the product form is a collection of the integrals of Fresnel type which are generally oscillatory~\cite{InomataEtAll1992}. A trick suggested by Feynman was to add a negative imaginary part to Planck's constant. This converts the oscillatory integrals into the Gaussian integrals and makes the path integral convergent.
 Other more generally applicable trick is to assume that each element of the diagonal mass matrix has a positive imaginary part~\cite{InomataEtAll1992}. Under this assumption, the path integral can be convergent independent of the metric of space.

 Natural trick is based on consecutive unfolding the Bohmian quantum corrector.
 The path integral computation stems directly from decomposition of
 the Schr{\"o}dinger equation to modified HJ equation plus the entropy balance equation.
 The Bohmian quantum corrector resulted from this decomposition
 expands the state space $S^{\,N}$ to imaginary sector. In turn,
 imaginary terms emergent in this computations suppress the wilder contributions to the path integral.
 Thus, we have non-trivial $N$-dimensional manifold embedded in the
 $2N$-dimensional complex state space, $\Sc^{\,N}$, where its real part is the conventional coordinate state space $S^{\,N}$.

% Section6
\section{\label{sec:level6}Concluding remarks.}

\begin{figure*}
  \begin{picture}(400,300)(40,15)
  \centering
  \includegraphics[scale=1]{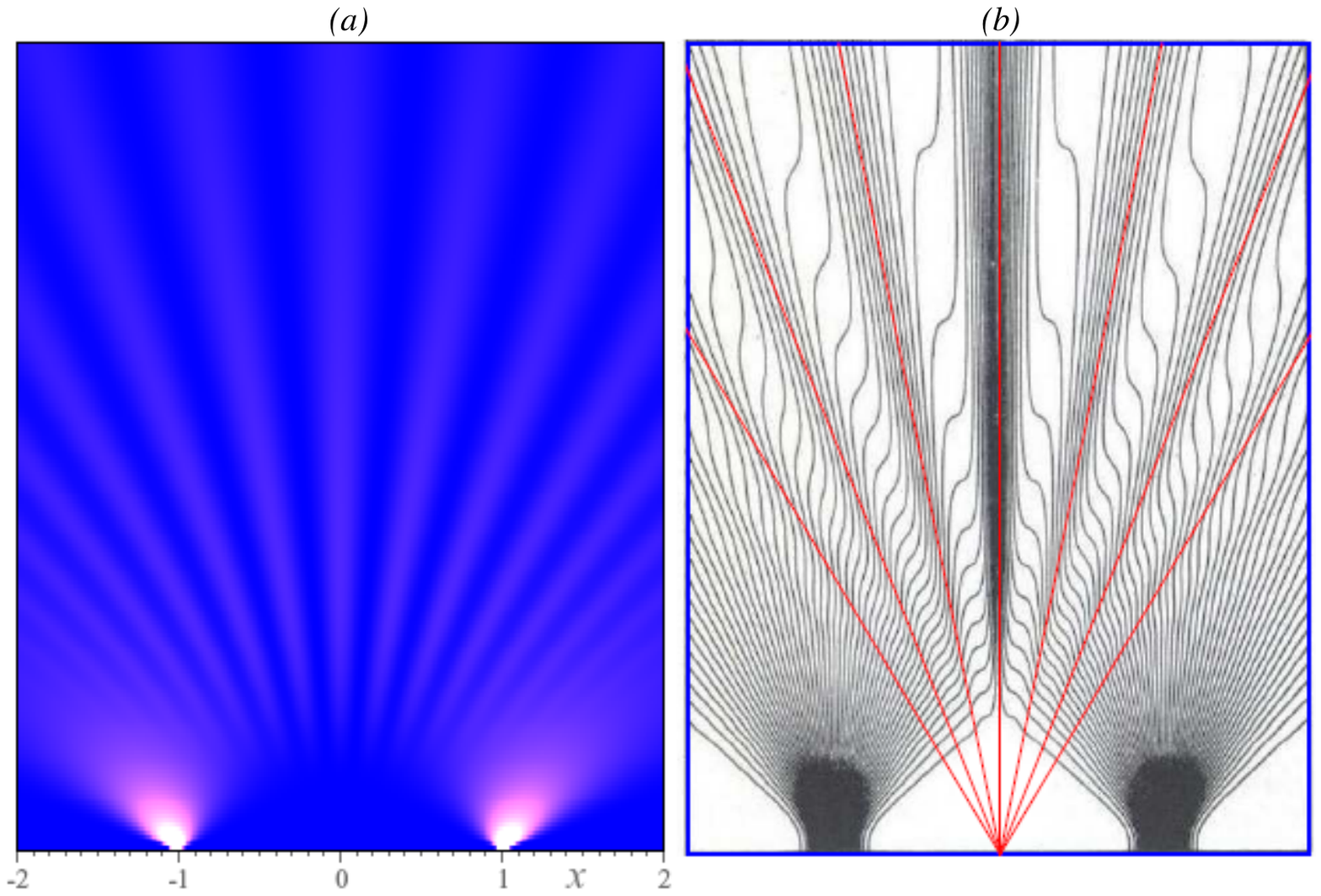}
  \end{picture}
  \caption{
  Two-slit interference experiment:
  (a) reconstructed scene of the two-slit interference;
  (b) two-slit interference map of the Bohmian trajectories.
  The black-white Bohmian trajectory map is shown in the Bohm's article~\cite{Bohm1990}.
  Red lines drawn in this map approximate the Bohmian trajectory beams.
    }
  \label{fig=3}
\end{figure*}

 Figure~\ref{fig=3}(a) shows a reconstructed scene of the two-slit interference experiment.
 Comparison with the two-slit interference maps of Bohmian trajectories  shown, for example, in articles~\cite{Bohm1990} and~\cite{CoffeyEtAll2008}
 reveals a qualitative agreement, see Figure~\ref{fig=3}(b).
 Red lines drawn in this figure approximate the Bohmian trajectory beams.
 They demonstrate a good agreement with the magenta rays clear visible in the left pattern, Figure~\ref{fig=3}(a).
 Bohmian trajectories are seen to be {\it geodesic trajectories} of an incompressible fluid loaded by the quantum potential~\cite{Wyatt2005}.

 Bohmian trajectories are trajectories submitted to the principle of least action that expands on the action integral~(\ref{qp=35})
 containing the complexified Lagrangian function. In fact, they stem from the complexified Euler-Lagrange, Hamilton-Jacobi mechanics.
 This complexified mechanics differs from the classical mechanics cardinally.
 For comparison, main formulas of the classical and complexified mechanics are collected in Tables~1 and~2 printed in pages~\pageref{Table1} and~\pageref{Table2}, respectively.
 Qualitative difference is extension of the formulas to the imaginary sector on a depth of order of smallness about $\hbar/m$.

 Expansion of three-dimensional coordinate space onto the imaginary sector
 underlying the complexified Lagrangian mechanics introduces new quality in evolution of quantum objects.
 One way to envisage the complex space is to imagine a hose-pipe.
 From a long distance it looks like a one dimensional line but a
 closer inspection reveals that every point on the line is in fact a {\it circle}.
 Such an expansion is not something strange.
 Currently, extra dimensions have become an accepted part of modern theoretical physics.

 It may initiate many speculations ranging from quantum state teleportation~\cite{BrownHiley2004} to
 backpropagation through time~\cite{Werbos2008b},~\cite{WerbosDolmatova2000}.
 It refers also to the fascinating Everett's "many-worlds" theory~\cite{DeWitt1973},~\cite{Everett1957}:
 the three dimensional universe or "world" that we see in everyday life is only one of the many "worlds"
 which exist side by side~\cite{Werbos2008a}.

\begin{acknowledgments}
 The author thanks Prof. Bill Poirier for supporting this work.
\end{acknowledgments}

% -----------------------------------------------------------

%\bibliography{bibtex}

\begin{thebibliography}{10}

\bibitem{Arnold1978}
V.~I. Arnold, \emph{Mathematical methods of classical mechanics}, (Springer, N.
  Y., 1978).

\bibitem{Bellman1957}
R.~Bellman, \emph{Dynammic programming}, (Princeton University Press,
  Princeton, N. J., 1957).

\bibitem{Bittner2000}
E.~R. Bittner, \emph{Quantum tunneling dynamics using hydrodynamic
  trajectories}, 
\urlprefix\url{http://arXiv.org/abs/quant-ph/0001119}, (18 Feb 2000).
% (arXiv.org/abs/quant-ph/0001119v2, 18 Feb 2000).


\bibitem{Bittner2003}
\bysame, \emph{Quantum initial value representations using approximate
  {B}ohmian trajectories}, 
 \urlprefix\url{http://arXiv.org/abs/quant-ph/0304012}, (2 Apr 2003).

\bibitem{Bohm:1952a}
D.~Bohm, \emph{A suggested interpretation of the quantum theory in terms of
  "hidden variables", {I}}, Phys. Rev. \textbf{85} (1952), 166--179.

\bibitem{Bohm:1952b}
\bysame, \emph{A suggested interpretation of the quantum theory in terms of
  "hidden variables", {II}}, Phys. Rev. \textbf{85} (1952), 180--193.

\bibitem{Bohm1990}
\bysame, \emph{A new theory of the relationship of mind and matter},
  Philosophical Psychology \textbf{3} (1990), no.~2, 271--286:
  \urlprefix\url{http://evans-experientialism.freewebspace.com/bohmphysics.htm}.
% \urlprefix\url{http://members.aol.com/Mszlazak/BOHM.html}.

\bibitem{BohmHiley1993}
D.~Bohm and B.~J. Hiley, \emph{The undivided universe: an ontological
  interpretation of quantum theory}, (Routledge, London, 1993).

\bibitem{Brillouin2004}
L.~Brillouin, \emph{Science and information theory}, (Courier Dover Publ.,
  Inc., N. Y., 2004).

\bibitem{BrownHiley2004}
M.~R. Brown and B.~J. Hiley, \emph{Schr{\"o}dinger revisited: an algebraic
  approach}, 
  \urlprefix\url{http://arXiv.org/abs/quant-ph/0005026}, (19~Jul 2004).

\bibitem{ChouWyatt2007}
{Chia-Chun Chou} and R.~E. Wyatt, \emph{Quantum trajectories in complex space},
  Phys. Rev. A \textbf{76} (2007), no.~012115.

\bibitem{CoffeyEtAll2008}
T.~M. Coffey, R.~E. Wyatt, and W.~C. Schieve, \emph{Monte {C}arlo generation of
  {B}ohmian trajectories}, 
  \urlprefix\url{http://arXiv.org/abs/0807.0209}, (1 Jul 2008).

\bibitem{DeWitt1957}
B.~S. DeWitt, \emph{Dynamical theory in curved spaces. {I. A} review of the
  classical and quantum action principles}, Rev. Mod. Phys. \textbf{29} (1957),
  377.

\bibitem{DeWitt1973}
B.~S. DeWitt and N.~Graham, \emph{The many-worlds interpretation of quantum
  mechanics}, (Princeton University Press, Princeton, {1973),~Contains Hugh
  Everett's original article.}

\bibitem{Dirac1933}
P.~A.~M. Dirac, \emph{The {L}agrangian in quantum mechanics}, Physikalische
  Zeitschrift der Sowjetunion \textbf{3} (1933), 64--72.

\bibitem{Dirac1945}
\bysame, \emph{On the analogy between classical and quantum mechanics}, Rev.
  Mod. Phys. \textbf{17} (1945), no.~2 and 3, 195--199.

\bibitem{Everett1957}
H.~Everett, \emph{Relative state formulation of quantum mechanics}, Rev. Mod.
  Phys. \textbf{29} (1957), 454--462.

\bibitem{FerrariStenger2004}
S.~Ferrari and R.~F. Stenger, \emph{Model-based adaptive critic designs}, in:
  Handbook of learning and approximate dynamic programming (J.~Si, A.~G. Barto,
  W.~B. Powell, and D.~Wunsch {II}, eds.), Wiley-IEEE, 2004, pp.~65--95.

\bibitem{Feynman1948}
R.~P. Feynman, \emph{Space-time approach to non-relativistic quantum
  mechanics}, Rev. Mod. Phys. \textbf{20} (1948), 367--387.

\bibitem{FeynmanHibbs1965}
R.~P. Feynman and A.~Hibbs, \emph{Quantum mechanics and path integrals},
  (McGraw Hill, N.~Y., 1965).

\bibitem{Grosche1993}
C.~Grosche, \emph{An introduction into the {F}eynman path integral},
  \urlprefix\url{http://arXiv.org/abs/hep-th/9302097}, (20 Feb 1993).

\bibitem{Grosche1996}
\bysame, \emph{Path integrals, hyperbolic spaces, and {S}elberg trace
  formulae}, (World Scientific, Singapore, 1996).

\bibitem{Hiley2002}
B.~J. Hiley, \emph{From the {H}eisenberg picture to {B}ohm: a new perspective
  on active information and its relation to {S}hannon information}, {in:
  Quantum Theory: reconsideration of foundations Proc. Int. Conf.} (Sweden)
  (A.~Khrennikov, ed.), (V{\"a}xj{\"o} University Press, June 2001) 2002,
  pp.~1--24.

\bibitem{InomataEtAll1992}
A.~Inomata, H.~Kuratsuji, and C.~C. Gerry, \emph{Path integrals and coherent
  states of {SU(2)} and {SU(1,1)}}, (World Scientific, Singapore, 1992).

\bibitem{JohnsonLapidus2002}
G.~W. Johnson and M.~L. Lapidus, \emph{The {F}eynman integral and {F}eynman's
  operational calculus}, (Oxford Science Publ., Oxford, 2002).

\bibitem{Lanczos1970}
C.~Lanczos, \emph{The variational principles of mechanics}, (Dover Publ., Inc.,
  N. Y., 1970).

\bibitem{LaValle2006}
S.~M. LaValle, \emph{Planning algorithms}, (Cambridge University Press,
  Cambridge, 2006).

\bibitem{MacKenzie2000}
R.~MacKenzie, \emph{Path integral methods and applications},
  \urlprefix\url{http://arXiv.org/abs/quant-ph/0004090}, (24~Apr 2000).

\bibitem{Morozov2005}
V.~B. Morozov, \emph{Electron}, 
 \urlprefix\url{http://phorum.lebedev.ru/viewtopic.php?t=14}
  (\textbf{6}, 2005).

\bibitem{Poirier2008c}
B.~Poirier, \emph{On flux continuity and probability conservation in
  complexified {B}ohmian mechanics}, 
 \urlprefix\url{http://arXiv.org/abs/0803.0193}, (3 Mar 2008).

\bibitem{Poirier2008a}
\bysame, \emph{Reconciling semiclassical and {B}ohmian mechanics: {I}.
  {S}tationary states}, 
 \urlprefix\url{http://arXiv.org/abs/0802.3472}, (23 Feb 2008).

\bibitem{Poirier2008b}
\bysame, \emph{Reconciling semiclassical and {B}ohmian mechanics: {V}.
  {W}avepacket dynamics}, 
 \urlprefix\url{http://arXiv.org/abs/0803.0143}, (2 Mar 2008).

\bibitem{PoirierParlant2008}
B.~Poirier and G.~Parlant, \emph{Reconciling semiclassical and {B}ohmian
  mechanics: {IV}. {M}ultisurface dynamics}, 
 \urlprefix\url{http://arXiv.org/abs/0803.0142}, (2 Mar 2008).

\bibitem{Poluyan2005}
P.~V. Poluyan, \emph{Nonclassical ontology and nonclassical movement},
  Kvantovaya Magiya \textbf{2} (2005), no.~3,~ 3119-3134:
  \urlprefix\url{http://quantmagic.narod.ru/volumes/VOL232005/p3119.html}.

\bibitem{Sargent1987}
T.~J. Sargent, \emph{Dynamic macroeconominc theory}, (Harvard University Press,
  Cambridge, Massachusetts, London, 1987).

\bibitem{Sbitnev2008a}
V.~I. Sbitnev, \emph{Bohmian split of the {S}chr{\"o}dinger equation onto two
  equations describing evolution of real functions}, Kvantovaya Magiya
  \textbf{5} (2008), no.~1,~ 1101-1111: 
 \urlprefix\url{http://quantmagic.narod.ru/volumes/VOL512008/p1101.html}.

\bibitem{Schr1926}
E.~Schr{\"o}dinger, \emph{An undulatory theory of the mechanics of atoms and
  molecules}, Phys. Rev. \textbf{28} (1926), no.~6, 1049--1070.

\bibitem{Titchmarsh1976}
E.~C. Titchmarsh, \emph{The theory of functions}, (Oxford Science Publ.,
  Oxford, {1976)}.

\bibitem{TrahanPoirier2008b}
C.~Trahan and B.~Poirier, \emph{Reconciling semiclassical and {B}ohmian
  mechanics: {II}. scattering states for discontinuous potentials},
  \urlprefix\url{http://arXiv.org/abs/0802.4069}, (27 Feb 2008).

\bibitem{TrahanPoirier2008c}
\bysame, \emph{Reconciling semiclassical and {B}ohmian mechanics: {III}.
  scattering states for discontinuous potentials}, 
 \urlprefix\url{http://arXiv.org/abs/0802.4053}, (27 Feb 2008).

\bibitem{Weigel1986}
F.~W. Weigel, \emph{Introduction to path-integral methods in physics and
  polymer science}, (World Scientific, Singapore, 1986).

\bibitem{Werbos2008a}
P.~J. Werbos, \emph{Bell's theorem, many worlds and backwards-time physics: Not
  just a matter of interpretation}, 
 \urlprefix\url{http://arXiv.org/abs/0801.1234}, (25 Mar 2008).

\bibitem{Werbos2008b}
\bysame, \emph{Specification of the q hypothesis: An alternative mathematical
  foundation for physics}, 
 \urlprefix\url{http://arXiv.org/abs/quant-ph/0607096}, (25 Apr 2008).

\bibitem{WerbosDolmatova2000}
P.~J. Werbos and L.~Dolmatova, \emph{The backwards-time interpretation of
  quantum mechanics - revisited with experiment}, 
 \urlprefix\url{http://arXiv.org/abs/quant-ph/0008036}, (7~Aug 2000).

\bibitem{Wyatt2005}
R.~E. Wyatt, \emph{Quantum dynamics with trajectories: Introduction to quantum
  hydrodynamics}, (Springer, N. Y., 2005).

\bibitem{WyattBittner2003}
R.~E. Wyatt and E.~R. Bittner, \emph{Quantum mechanics with trajectories:
  Quantum trajectories and adaptive grids},
  \urlprefix\url{http://arXiv.org/abs/quant-ph/0302088}, (11 Feb 2003).

\end{thebibliography}
\providecommand{\bysame}{\leavevmode\hbox to3em{\hrulefill}\thinspace}
\providecommand{\MR}{\relax\ifhmode\unskip\space\fi MR }
\providecommand{\MRhref}[2]{%
  \href{http://www.ams.org/mathscinet-getitem?mr=#1}{#2}
}
%\expandafter\ifx\csname urlprefix\endcsname\relax\def\urlprefix{URL }\fi

\providecommand{\href}[2]{#2}

% -----------------------------------------------------------

\end{document}